\documentclass[sigconf,natbib=false,anonymous=false]{acmart}
\AtBeginDocument{%
  }

\setcopyright{acmlicensed}
\copyrightyear{2026}
\acmYear{2026}
\acmDOI{XXXXXXX.XXXXXXX}
\acmConference[SIGIR Workshop AgentSearch]{}{July 24,
  2026}{Melbourne, Australia}
\acmISBN{978-1-4503-XXXX-X/2018/06}

\RequirePackage[
  datamodel=acmdatamodel,
  style=acmnumeric,sorting=none,
  ]{biblatex}

\begin{document}

\title{LLM Retrieval for Stable and Predictable Ad Recommendations}

\author{
  Vinodh Kumar Sunkara,
  Satheeshkumar Karuppusamy,
  Hangjun Xu,
  Sai Deepika Regani,
  Kshitij Gupta,
  Gaby Nahum,
  Sneha Iyer,
  Jean-Baptiste Fiot,
  Yinglong Guo,
  Xiaowen Guo,
  Atul Jangra,
  Yucheng Liu,
  Jinghao Yan,
  Vijay Pappu,
  Benjamin Schulte,
  Deepak Chandra
}

\affiliation{
  \institution{Meta Platforms, Inc.}
  \city{Menlo Park and New York}
  \country{USA}
}

\email{
  vinodhsunkara, satheeshk, hangjunxu, deepikaregani, kshitijgupta, gnahum12345, snehaiyer, jbfiot, yinglongguo ( @meta.com)
}
\email{
  xiaoweng, atuljangra, yuchengliu, jyan, psnvijay, bschulte, deepakchandra ( @meta.com)
}

\begin{abstract}
Traditional ads recommendation systems have primarily focused on optimizing for prediction accuracy of click or conversion events using canonical metrics such as recall or normalized discounted cumulative gain (NDCG). With the hyper-growth of ads inventory and liquidity with generative AI technologies, the prediction \emph{stability} and \emph{predictability} is becoming increasingly critical. Intuitively, prediction stability and predictability can be defined to quantify system robustness with respect to minor/noisy input (ads, creatives) perturbations, the lack of which could lead to advertiser perceivable problems such as \emph{repeatability}, \emph{cold start} and \emph{under-exploration}. In this paper, we introduce a new evaluation framework for quantifying stability and predictability of an ads recommender system, and present an online validated semantic candidate generation framework powered by fine-tuned Large Language Models (LLMs) that showed significant improvement along these metrics by fundamentally improving the semantic-awareness of the system. The approach extracts hierarchical semantic attributes from ad creatives to obtain LLM representations, which serve as the foundation for graph-based expansion, ensuring the retrieved candidates encapsulate semantic variants of an ad, guaranteeing that small creative variants from the advertiser yield consistent and explainable delivery results to the user. We tested this LLM ads retrieval framework in a large-scale industrial ads recommendation system, demonstrating significant improvements across offline and online A/B experiments, showcasing gains in both predictability and traditional performance metrics. Although evaluated in the ads stack, this is a general framework that can be applied broadly to any large-scale recommendation and retrieval systems facing similar scaling and predictability challenges.
\end{abstract}



\maketitle

\section{Introduction} \label{introduction}
In the rapidly evolving ads recommendation landscape, personalized recommendations \cite{naumov2019dlrm} have been effective in enhancing user experiences and ad performance. As the complexity and volume of ads increase, it becomes more important to develop deeper semantic understandings and be able to perform the computation at different semantic granularities based on the resources and task requirements. With this scaling, \emph{predictability} of system outputs becomes crucial. Traditional evaluation metrics focus on prediction accuracy and recall (e.g., normalized entropy, recall, NDCG), but there is little industry focus on quantifying or optimizing for predictability, especially in ads recommendations. 

A predictable system is one whose outputs change in a \emph{robust}, \emph{explainable}, and \emph{causal} manner in response to input, data, or parameter changes aligning with advertiser expectations. We introduce new metric to quantify the \emph{predictability}. The problem has been addressed in ranking stage by moving the features from raw ad IDs to semantic IDs \cite{2025SemanticIdPaper}. This paper focuses on the retrieval stage leveraging Large Language Models (LLMs) for semantically aware candidate generation.

Recent advances in retrieval-augmented Large Language Models (LLMs) have transformed the landscape of recommendation systems, enabling more robust semantic matching and context-aware candidate generation \cite{kim2024large,lyu2023llm,vats2024exploring}. Unlike traditional models that rely heavily on sparse features or manual heuristics, LLM-based retrieval leverages deep contextual embeddings to capture nuanced semantic understanding and reasoning of ad entities for appropriate user targeting. This approach not only improves the precision of candidate selection but also enhances the system’s predictability across diverse creative variants and campaign objectives. By integrating LLM retrieval into the candidate generation stage, recommendation platforms can achieve greater stability, explainability, and adaptability in ad delivery, addressing longstanding challenges in scaling and predictability for large-scale recommendation systems \cite{deldjoo2024recommendation,zhang2023recommendationinstructionfollowinglarge,yang2023palrpersonalizationawarellms}.

The paper is organized as follows: Section [\ref{metric_definition}] defines the metrics used to measure the predictability and performance in the ads delivery system to showcase it's improvement from the proposed LLM solution. Section [\ref{solution:LLM overview}] presents the high-level design of using fine-tuned LLMs \cite{touvron2023llama} for semantic ad attributes and ad-to-ad similarity-based candidate generation. Section [\ref{solution:LLM system}] covers engineering considerations for the end-to-end LLM candidate generation system. Section [\ref{solution:evaluation}] presents evaluation metrics and shows improved predictability. Section [\ref{future work}] summarizes contributions and future directions for generalizability (A/B).

\section{Metric Definitions} \label{metric_definition}
As introduced in Section [\ref{introduction}], we establish a framework for measuring system predictability and performance enhancements from the LLM Retrieval solution leveraging the below suite of metrics. 
\begin{itemize}
\item{\texttt{Recall@k}}: Recall@K measures set-level relevance and is widely used for retrieval evaluation.
\item{\texttt{Online top-line metric}}: We report the online top-line metric within a large-scale industrial ads recommendation system to measure delivered ads value in online A/B experiments.
\item {\texttt{Predictability (a.k.a. A/A' difference)}}: Maintain consistent delivery performance over time, even with minor changes in ad inputs (hence the term A/A', where A' represents a minor perturbation of A, where A is the input entity such as an ad or a creative). It ensures ad performance doesn’t fluctuate unpredictably or diverge drastically with small updates to ad creatives preventing delivery inefficiencies.
\item{\texttt{Median Absolute Deviation (MAD)}}: This metric quantifies the variability of the above defined A/A’ difference across the time period. 
\end{itemize}

We discuss in more depth about the construction of the predictability metric since it is a relatively new concept. On a high level, it is measuring the output differences of the system with respect to input perturbations (i.e. A vs A') without fundamental change of the semantic meaning. To define this system level A/A' difference, we built an internal framework that publishes a copy (called the "shadow ad") of the original ad (called the "primary ad"). We measure predictability (i.e. the A/A' differences) by comparing normalized conversions between original ('primary') ads and their copies ('shadow' ads).

For notation, for a given (primary, shadow) ads pair $(ad^p, ad^s)$, we denote their relative difference of their conversions to be $\Delta$. Under the hypothesis that these two ads are delivered equally, a Gaussian  approximation of the difference (or variance) of $\Delta$ is ${2/}{(conv(ad^p) + conv(ad^s))}$, where $conv(\cdot)$ stands for the number of optimized conversions of the ad.  We define the Stat-Sig Difference between $ad^p$ and $ad^s$, denoted as \text{StatSigDiff}($ad^p$, $ad^s$), by:
\begin{small}
\begin{equation}
\text{StatSigDiff}(ad^p, ad^s) = \max\Big(0, \Delta - 1.65 * \sqrt{\frac{2}{conv(ad^p) + conv(ad^s)}}\Big).
\label{eqn:definition_AA}
\end{equation}
\end{small}
Here the $1.65$ standard deviation is chosen to represent the 90\% confidence interval under a Gaussian distribution at null hypothesis. And the system level $\textbf{StatSigDiff}$ metric would just be the aggregated sum of \text{StatSigDiff}($ad^p$, $ad^s$) over all (primary, shadow) pairs $(ad^p, ad^s)$ normalized by the total revenue:

\begin{small}
\begin{equation}
\text{StatSigDiff} = \sum_{i = 1}^{N} \frac{\text{StatSigDiff}(ad_i^p, ad_i^s) * \sqrt{rev(ad_i^p) + rev(ad_i^s)}}{\sum_{i = 1}^{N} \sqrt{rev(ad_i^p) + rev(ad_i^s)}},
\label{eqn:statsigdiff_aggr}
\end{equation}
\end{small}
where $N$ is the total number of (primary, shadow) pairs we generate, and $rev(\cdot)$ is the total revenue of the ad.

For A/A', the shadow is a copy with a unique ad ID along with a key feature perturbed to simulate a minor change. A low StatSigDiff reflects strong semantic awareness, while a high value indicates reliance on non-semantic features. StatSigDiff, though not expected to be zero due to inherent randomness, serves as a critical guardrail for model development.

\section{LLM Based Ad Candidate Generation} \label{solution:LLM overview}
Most large scale ads recommendation systems use a cascading multi-stage paradigm with multi-routed \cite{youtube_recommendations} candidate generation stage followed by one or more re-ranking stage(s). The key to improve the predictability of the system (defined in Equation [\ref{eqn:statsigdiff_aggr}]) is to find the right semantic representation of the user and the items. For the scope of this paper, we focus on the candidate generation stage. The primary objective of a candidate generator is to increase the final stage recall, passing as many potentially relevant candidates to the next stage as allowed by infra capacity.  Given that it is ultimately the semantic concepts (product being sold, apps being promoted) that the users engage with, a good representation should be \emph{deeply semantic} that can capture these high level concepts. A \emph{hierarchical structured} representation would be useful to facilitate a hierarchical, cluster based computational framework for an efficient processing.  These two properties, semantic-aware and hierarchical, are thus the key properties we seek. And large-language models (LLM) such as Llama is thus a good candidate model due to their:

\textbf{Rich Contextual Understanding}: LLMs process vast contextual information \cite{llm_recommendations}, enabling nuanced ad representations and more accurate candidate selection.

\textbf{Semantic Awareness}: LLMs leverage ad inputs (title, description, etc.) to capture semantic meaning, grouping similar ads closely in semantic space.

\textbf{Hierarchy/Graph Structure from LLM metadata}: LLM generated metadata enables building a graph of ads, where nodes are ads and edges represent shared attributes, supporting efficient candidate expansion and system predictability.

Our high level design to leverage LLM in ads candidate generation to improve system predictability (A/A' difference) is the following: 
\begin{itemize}
    \item We take a pre-trained LLM, fine-tune it on ads data set with retrieval specific tasks. In our case we already have an in-house fine-tuned LLM on ads engagement data sets.
    \item Use the LLM to infer hierarchical ad semantic attributes and a relationship graph based on the attribute similarity (e.g., Jaccard similarity).
    \item Apply graph traversal algorithms to expand from seed ads, enabling semantic ad-to-ad similarity expansion and optimizing recall.
\end{itemize}

The fundamental hypothesis is that the fine-tuned LLM would be able to better retrieve ad candidates that belong to the same semantic equivalence class that would otherwise been missed.

In the following Section [\ref{solution:LLM system}], we present the end-to-end system design and the above core components. 

\section{End-to-End System Design} \label{solution:LLM system}
There are four main system components that implement the LLM based ads candidate generation:

\subsection{LLM-Driven Ads Representation Learning}
The process of representation learning for ads involves transforming ad creatives and product descriptions into high-dimensional semantic vectors. This transformation is achieved using advanced Instruct Large Language models, which encode world knowledge into retrieval specific discrete tokens, and then construct a semantic graph based on them, enabling the retrieval of relevant candidates for ad-to-ad expansion. This approach captures detailed textual and categorical nuances, significantly improving the quality of subsequent recommendations.

\subsection{Scalable Infrastructure for LLM Processing}
\begin{figure}[h]
  \centering
  \includegraphics[width=\linewidth]{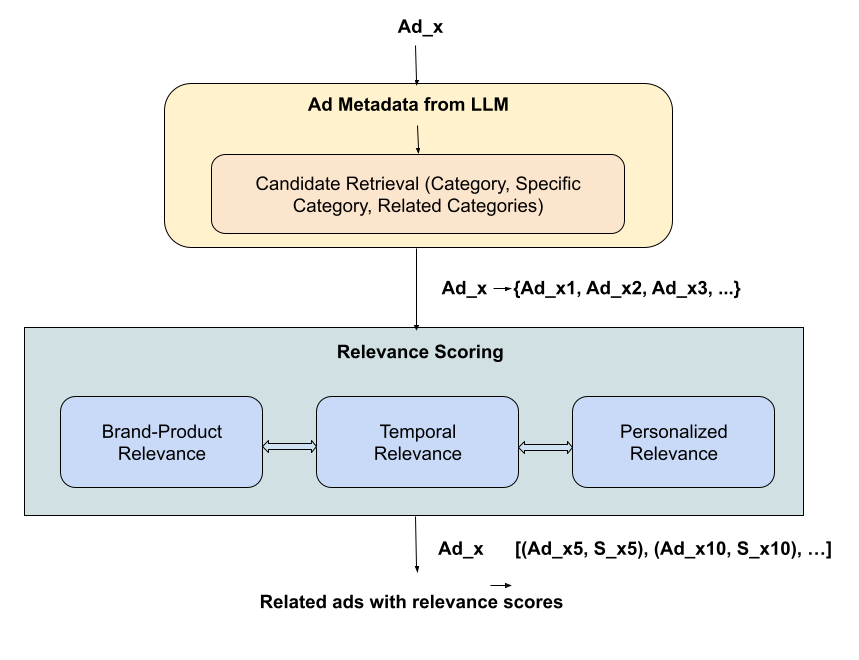}
  \caption{LLM Ad to Ad Generator}
  \Description{Ad creatives to category, attribution generation using LLM}
  \label{ad_to_ad_pipeline}
\end{figure}

A robust and scalable infrastructure is required for LLM processing, ensuring the generation of ad-level representations at scale. The infrastructure includes a distributed processing framework that meets stringent latency and throughput demands. As shown in Figure~\ref{ad_to_ad_pipeline}, key components include LLM ad metadata generation, Ad to Ad Relevance Scoring, and horizontal scaling across high-performance GPUs. This setup facilitates efficient batch inference and boosts throughput, enabling the system to handle large volumes of data effectively. 

\subsection{Semantic Navigation - Graph Traversal Algorithms}
\begin{figure}[h]
  \centering
  \includegraphics[width=\linewidth]{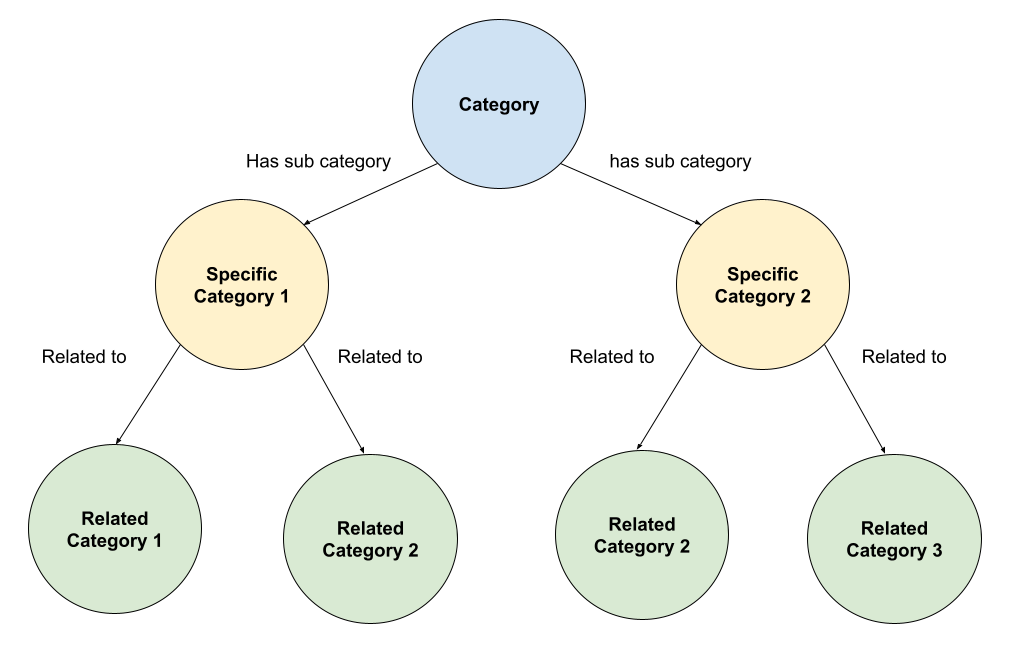}
  \caption{LLM Contextual Category Graph}
  \Description{LLM Contextual Category Graph}
  \label{contextual_graph}
\end{figure}

Semantic navigation uses a graph built from LLM-extracted categories and attributes to expand candidate ads. By traversing this semantic graph, the system finds clusters of contextually similar ads, enabling dynamic and relevant recommendations (see Figure~\ref{contextual_graph}).

\begin{itemize}
\item {\texttt{Step1 - Retrieval}}: Retrieval step focuses on retrieving approximately relevant candidates to every candidate by computing categorical relevance using LLM Representations.

\item {\texttt{Step2 - Relevance Scoring}}: Relevance scoring step focuses on scoring ad to ad similarity based on more deeper dimensions like brand, product and contextual attributes
\end{itemize}

For both steps, fuzzy set matching using Jaccard similarity is applied, starting with phrase matches and backing off to token matches

\[
S_{\text{R}}(Ad1, Ad2) = 
\begin{cases}
S(P_{Ad1}, P_{Ad2}) & \text{if } S(P_{Ad1}, P_{Ad2}) \geq \theta \\
S(T_{Ad1}, T_{Ad2}) & \text{otherwise}
\end{cases}
\]
\[
\text{where } P_{Ad1}, P_{Ad2} \text{ are sets of phrases, and }
\text T_{Ad1}, T_{Ad2} \text { are sets of tokens.}
\]

\subsection{Real-Time Candidate Retrieval and Service}
The real-time candidate retrieval and service layer offers high-throughput retrieval and compatibility with downstream ranking modules. The real-time candidate retrieval service framework we built serves as the foundational infrastructure for candidate generation, with optimized LLM inference to maximize utilization of the GPU hosts. This setup ensures low-latency service and efficient online retrieval, supporting the delivery of personalized and context-aware recommendations.

Leveraging the four foundational pillars above, the LLM candidate generation primarily happens in the below two stages as shown in the Figure~\ref{ad_to_ad_pipeline}:

\subsubsection*{\textbf{Stage 1: Categories, Attributes and Contextual Caption Generation}}

\paragraph{Ad to Categories:} 
\begin{equation}
    \quad f_{1}(Ad) \longrightarrow S_{text{Ad}(C)} \longrightarrow \{(c_1, s_1), (c_2, s_2), \cdots, (c_n, s_n)\},
    \small
\end{equation} where $S_{\text{Ad}(C)}$ is the categories for the ad, $C$ is the category and $s$ is the score. 

\subsubsection*{\textbf{Stage 2: Similar Ads Retrieval}}

\paragraph{Retrieval function:}
\begin{equation}
    S_{R}(Ad1, Ad2) = \sum (S_{Ad1(c)} \times S_{Ad2(c)}), \ \text{for} \ c \in C_{Ad1} \cap C_{Ad2} 
    \small
\end{equation}

\section{Experiment Design and Evaluation} \label{solution:evaluation}
\subsection{Infra setup}
For evaluation, we employed the open-source text-only LLama3-8B Instruct model \cite{llama3herd2024}, leveraging its zero-shot inference capabilities. The data set comprises approximately tens of millions of data points and only textual descriptions were utilized to generate supplementary metadata. The evaluation is conducted as an online A/B test with setup as follows:
\begin{itemize}
\item {\texttt{Test}}: Test arm has an LLM-based candidate generator introduced at the ad retrieval stage while keeping the rest of the delivery flow unchanged. 
\item {\texttt{Baseline}}: An ensemble of candidate generators such as Two-tower, embedding and graph-based generators that serve as a baseline to demonstrate the incremental value of the LLM based semantic candidate generator.
\end{itemize}


\subsection{Evaluation framework}
For evaluation metrics, we look at 1) canonical ad performance metrics such as ad clicks and conversion, 2) A/A' predictability metric (defined in Equation [\ref{eqn:statsigdiff_aggr}]).  We also measure the \emph{median absolute deviation} (MAD) of the relative difference of impressions between (primary, shadow) ad pairs on a daily basis to quantify the variability of A/A’ difference in the time period:
$$
\text{rel. diff. of impression} := \frac{\sum_{(ad^p, ad^s)} \text{Impression of } ad^p }{\sum_{(ad^p, ad^s)} \text{Impression of } ad^s} - 100\%,
$$

and 
\[
MAD := median\left(|\text{rel. diff. of impression} (day_i) - m|\right).
\]

Here $m$ is the median of the relative difference of impressions across the testing days.

\subsection{Results}
\subsubsection{Delivery Performance Improvements}
For the canonical ads performance metrics, our experiment showed statistically significant $0.45\%$ lift in performance on the topline online metric for the experimentation segment. From a retrieval performance standpoint, we also see our treatment increased the final stage recall by $1.2\%$, validating the core hypothesis of our semantic candidate generation approach in capturing end user intent and delivering highly relevant content. 

\begin{table}[h]
\centering
\small
\begin{tabular}{ccc}
\toprule
\textbf{Top-K} & \textbf{Recall Alignment Ratio} & \textbf{Incremental Recall Potential} \\
\midrule
5   & \textbf{0.51X} & 1.00Y (baseline) \\
10  & \textbf{0.44X} & 1.15Y \\
50  & 0.21X          & 1.62Y \\
100 & 0.13X          & 1.77Y \\
200 & 0.07X          & \textbf{1.89Y} \\
\bottomrule
\end{tabular}
\caption{Delivery Performance Improvements}
\label{tab:llm_recall_metrics}
\end{table}

The LLM-based candidate generator demonstrates strong quality-aware ranking behavior, with Recall Alignment Ratio decreasing from 0.51X at Top-5 to 0.07X at Top-200, a 7 times higher concentration of quality alignment at smaller K values. As shown in the Table~\ref{tab:llm_recall_metrics}, assuming 1.0X = 100\% of Top-K candidates, this monotonic pattern validates that the LLM effectively prioritizes high-quality candidates aligned with the baseline system's top recommendations. Notably, even at Top-5, the LLM achieves an Incremental Recall Potential of 0.49X (say Y), expanding to 1.89 times at Top-200, demonstrating substantial capacity to augment existing retrieval systems with diverse, complementary recommendations.

\subsubsection{System Predictability (A/A') Improvements}
\begin{figure}[ht]
  \centering
  \includegraphics[width=\linewidth]{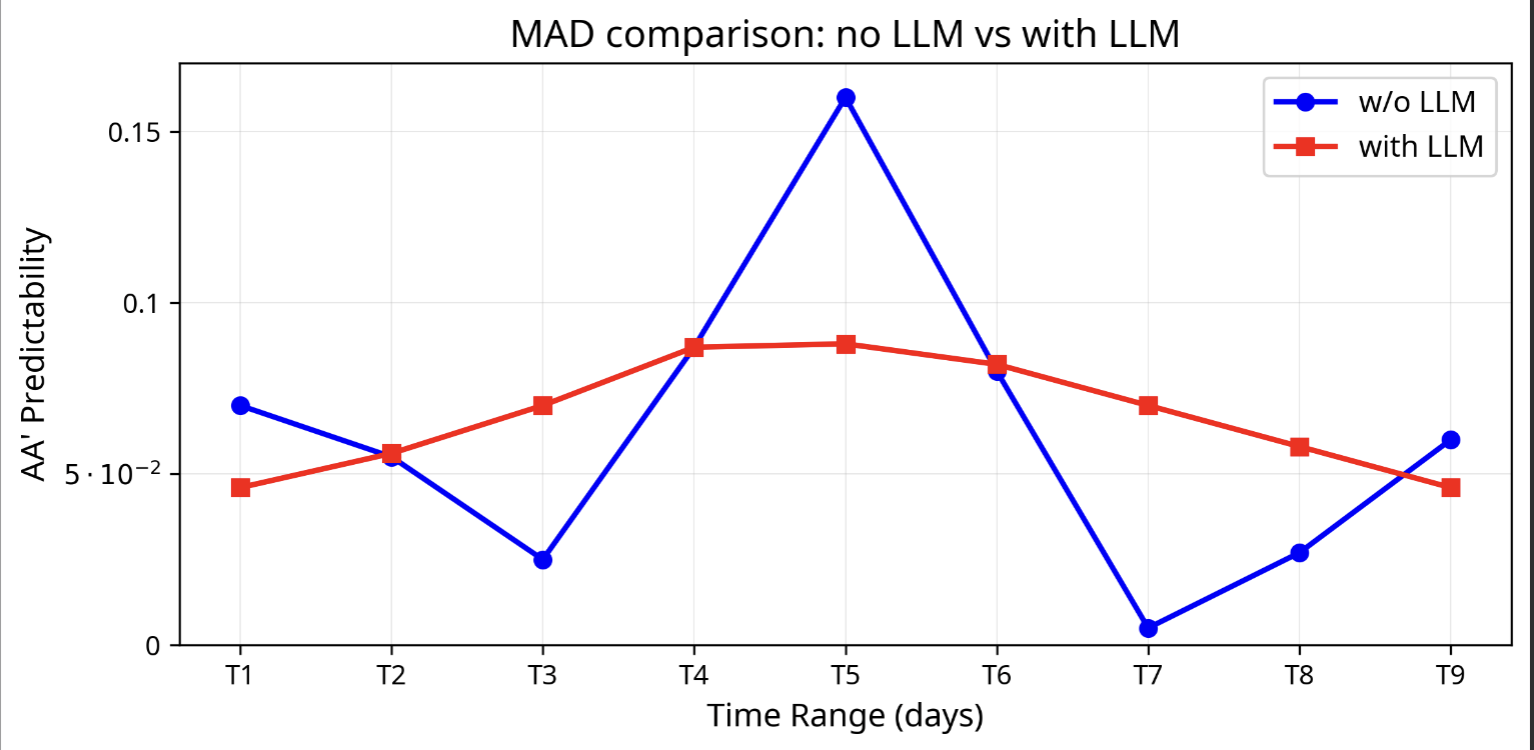}
  \caption{Daily impression relative difference between the A/A’ ad pairs for the test vs control.}
  \Description{Daily impression relative difference between the AA’ ad pairs for the test vs control.}
  \label{fig:MAD}
\end{figure}

The experiment demonstrated a statistically significant relative reduction in the top-line A/A’ difference of $8.62\%$ comparing test vs control in the online A/B test. Moreover, as shown in Figure~\ref{fig:MAD}, we observed a 45\% improvement in MAD of the daily impression difference in the test compared to the control version. This improvement underscores the effectiveness of our approach in reducing variance and achieving more consistent ad performance through the lens of the retrieval layer. By leveraging advanced language models and semantic candidate generation techniques, the proposed approach demonstrates the potential to deliver more personalized and context-aware recommendations in the initial stages of ad serving. These improvements, as evidenced by our experimental results, suggest greater alignment with advertiser objectives, increased predictability, and enhanced overall ad performance.

\section{Future Work} \label{future work}
Building on our work with LLM powered semantic candidate generation and new metric for ads recommendation predictability, our future focus is to develop an end-to-end, semantically aware ad recommender system with enhanced generalizability at every stage. We plan to leverage advanced, fine-tuned language models and integrate multimodal, real-time adaptive learning targeting generalizabilty and cold-start improvements. 
\printbibliography

\end{document}